\documentclass[a4paper,12pt]{article} 

\usepackage{amsmath}
\usepackage{color}
\usepackage{amssymb, amsfonts, amsthm, amscd}
\usepackage{bm}
\usepackage[dvips]{graphicx}
\usepackage{slashed}
\usepackage{url}
\usepackage[all]{xy}

\begin{document}

\thispagestyle{empty}
\vspace*{-15mm}
{\bf OCHA-PP-363}\

\vspace{12mm}

\begin{center}

{\Large\bf
 Nambu Dynamics\\ 
and Hydrodynamics of Granular Material}\footnote{Talk given at the workshop on ``Space-time topology behind formation of micro-macro magneto-vortical
structure manifested by Nambu mechanics'' held at Osaka City University, Sept.28-Oct.1 (2020)}

\vspace{7mm}

{\bf Akio Sugamoto$^{1}$, Kazuharu Bamba$^{2}$, Tetuya Kawamura$^{3}$, Anna Kuwana$^{4}$, Yusaku Nagata$^{5}$, Mayumi Saitou$^{1}$}
\vspace{2mm}

{\it
$^1$Department of Physics, Graduate School of Humanities and Sciences, \\
   Ochanomizu University, Tokyo 112-8610, Japan  \\
$^2$Division of Human Support System, Faculty of Symbiotic Systems Science, \\ Fukushima University, Fukushima 960-1296, Japan\\
$^3$Department of Informaton Scienes, Graduate School of Humanities and Sciences, \\
   Ochanomizu University, Tokyo 112-8610, Japan \\
$^4$Division of Electronics and Informatics, Faculty of Science and Technology, \\ Gunma University,
  1-5-1 Tenjin-cho, Kiryu City, Gunma 376-8515, Japan \\
$^5$School of Infomaon Science, \\
Meisei University, 2-1-1 Hodokubo, Hino 191-8506, Japan 
}\\

\end{center}

\vspace{10mm}
\begin{center}
\begin{minipage}{14cm}
\baselineskip 16pt
\noindent
\begin{abstract}
On the basis of the intimate relation between Nambu dynamics and the hydrodynamics,
the hydrodynamics on a non-commutative space (obtained by the quantization of space), proposed by Nambu in his last work, is formulated as ``hydrodynamics of granular material''.

In Part 1, the quantization of space is done by Moyal product, and the hydrodynamic simulation is performed for the so obtained two dimensional fluid, which flows inside a canal with an obstacle.  The obtained results differ between two cases in which the size of a fluid particle is zero and finite.  The difference seems to come from the behavior of vortices {\color{black}{generated by an obstacle}}.

In Part 2 of quantization, considering vortex as a string, two models are examined; one is the ``hybrid model'' in which vortices interact with each other by exchanging Kalb-Ramond fields (a generalization of stream functions), and the other is the more general ``string field theory'' in which Kalb-Ramond field is one of the excitation mode of string oscillations.  In the string field theory, Altarelli-Parisi type evolution equation is introduced.  It is expected to describe the response of distribution function of vortex inside a turbulence, when the energy scale is changed.  The behaviour of viscosity differs in the string theory, being compared with the particle theory, so that Landau theory of fluid to introduce viscosity may be modified.

In conclusion, the hydrodynamics and the string theory are almost identical theories.   It should be noted, however, that the string theory to reproduce a given hydrodynamics is not a usual string theory.

\end{abstract}

\end{minipage}
\end{center}

\newpage

 \section{Nambu Dynamics (1973)\cite{Nambu1973}}
Hamiltonian dynamics is given in the two dimensional phase space {\color{black}{$(x, p)$}}, with a single Hamiltonian $H$;
\begin{eqnarray}
\dot{x}=\frac{\partial H}{\partial p}=\{x, H\}, ~~\dot{p}=-\frac{\partial H}{\partial x}=
{\color{black}{\{p, H\}}}.
\end{eqnarray}

Nambu generalized this to D-dim phase space $(x_1, x_2, \cdots, x_D)$
with D$-$1 Hamiltonians $(H_1, H_2, \cdots, H_{D-1})$; 
\begin{eqnarray}
\dot{x}_i=\{ x_i, H_1, H_2, \cdots, H_{D-1} \}, ~{\color{black}{(i=1,\cdots, D)}}, \label{EOM}
\end{eqnarray}
where the r.h.s. is Jacobian, called ``Nambu bracket'':
\begin{eqnarray}
\{A_1, A_2, \cdots, A_D\}_{NB}= \frac{\partial (A_1, A_2, \cdots, A_D)}{\partial(x_1, x_2, \cdots, x_D)}.
\end{eqnarray}

Liouville theorem that the phase space volume is
unchanged gives {\color{black}{incompressible}} fluid. The peculiar ``advective term'', an origin of non-linearity in hydrodynamics, appears naturally in Nambu dynamics;
\begin{eqnarray}
v^i=\{x^i, {\color{black}{H_1}}, H_2, \cdots, H_{D-1}\}_{NB}=\sum_{j} \frac{\partial v_i}{\partial x_j} \{ x_j, H_1, H_2, \cdots, H_{D-1} \}, 
\end{eqnarray}
since it gives $\dot{v^i}=({\bf v} \cdot {\bf \nabla})v_i$.

Hamiltonians are equal to the stream functions in hydrodynamics $\{\psi_1, \psi_2, \cdots, \psi_{D-1}\}$
\begin{eqnarray}
v^i=\{x^i, \psi_1, \psi_2, \cdots, \psi_{D-1}\}_{NB}=\epsilon^{ij_1j_2\cdots j_{D-1}} \frac{\partial \psi_1}{\partial x^{j_1}} \cdots \frac{\partial \psi_{D-1}}{\partial x^{j_{d-1}}}
\end{eqnarray}
which implies $\bm{\nabla \cdot v}=0$, and the velocity points along the intersection line of all the stream functions be constant.

Navier-Stokes equation for incompressive fluid reads
\begin{eqnarray}
\rho \left(\dot{\bf v} - ({\bf v} \cdot {\bf \nabla}){\bf v}\right)= -{\bf \nabla} (p + V(x)) + \mu \Delta {\bf v},
\end{eqnarray}
where $p$ is pressure, $\mu$ is shear viscosity with
dimension [Pas][s]. We also use $\nu=\mu/\rho$ with dimension [$m^2$/s]. $V(x)$ is an external potential of force.

The r.h.s. (force term) can also be written by using Nambu
brackets:
\begin{eqnarray}
-{\bf \nabla}^i ~p =-\frac{1}{(D-1)!}\epsilon^{ijk \cdots}\{p, x_j, x_k, \cdots\}, \\
- \mu \Delta {\bf v}_i = \frac{-\mu}{(D-1)!} \{\{{\bf v}_i, x_j, x_k, \cdots \}, x_j, x_k, \cdots \}.
\end{eqnarray}

In his last work, Nambu proposed ``the quantization of
space" (2011, 2013), or the introduction of non-commutative space in
hydrodynamics \cite{Nambu(2011-2013)}. His intension is to understand
the empirical ``Bode law'', that is, the radius of the n-th planet follows $R_n=a \; 2^n+b$.  

We [SaBSu(2014)] interpreted his hydrodynamics as
``Hydrodynamics of Granular Material"\cite{SaBSu(2014)}.

\section{Quantization of Nambu's hydrodynamics (Part 1)}
To quantize the Nambu bracket, we can use the Moyal
product, familiar for non-commutative space.
For D=2, 3 hydrodynamics, the Moyal products are given by 
\begin{eqnarray}
&&(A \ast B) (x)= \left. e^{i \frac{\theta_2}{2!}  \frac{\epsilon_{ab}\partial^2}{\partial y^a\partial z^b}} A(y)B(z) \right|_{y, z \to x}, \\
&&(A \ast B \ast C) (x)= \left. e^{ i \frac{\theta_3}{3!} \frac{\epsilon_{abc}\partial^3}{\partial y^a\partial z^b\partial u^c}} A(y)B(z)C(u) \right|_{y, z, u \to x} .
\end{eqnarray}

Therefore, the quantization is carried out by replacing
\begin{eqnarray}
i \theta_2 {\color{black}{\{A, B\}}} \rightarrow [A(x), B(x)]_{M}= ~A(x) \ast B(x)-B(x) \ast A(x), \label{Moyal2}
\end{eqnarray}
and 
\begin{eqnarray}
i \theta_3 {\color{black}{\{A, B, C\}}} \rightarrow [A(x), B(x), C(x)]_{M}=  ~A(x) \ast B(x) \ast C(x)+ \cdots. \label{Moyal3}
\end{eqnarray}
The {\color{black}{Planck}} constant in our case is the minimum area $\theta_2$ in
2D and minimum volume $\theta_3$ in 3D of a granular particle which are {\color{black}{identified}} as the particle size.

Then, the additional force $K$ appears, as the finite size
effect or the quantum effect of fluid particle, 
\begin{eqnarray}
\hspace{-5mm}
K&=& \left. \frac{(\theta_2)^2}{24} \rho \left( \partial_{y_1}\partial_{z_2}-\partial_{y_2}\partial_{z_1}\right)^2 \sum_{a=1, 2} \partial_{y_a} \mbox{\boldmath $v$} (y) \mbox{\boldmath $v$}_a (z) \right|_{y, z \to x} 
\quad 
\mbox{(2D)}, \\
\hspace{-5mm}
K&=& \left. \frac{(\theta_3)^2}{3!} \rho \, \epsilon_{v, \psi_1, \psi_2} \left( \left( \frac{1}{3!}\sum_{abc} \frac{\partial^3}{\partial_{y_a}\partial_{z_b}\partial_{u_c} }\right)^3  \mbox{\boldmath $v$}(y) \psi_1(z)\psi_2(u)\right) \right|_{y, z, u \to x}
\quad 
\mbox{(3D)},
\end{eqnarray}
where $(\psi_1(x), \psi_2(x))$ are stream functions.

Next, we perform the hydrodynamic simulation in 2D [KaKuNaSaSu(2018)]\cite{KaKuNaSaSu(2018)}.

Fluid flows from the left to the right {\color{black}{inside a canal}}; with a table-shaped
obstacle {\color{black}{making a upper}} small and {\color{black}{a lower}} large slits.

{\begin{figure}[h]
\centering
\includegraphics[width=140mm,clip]{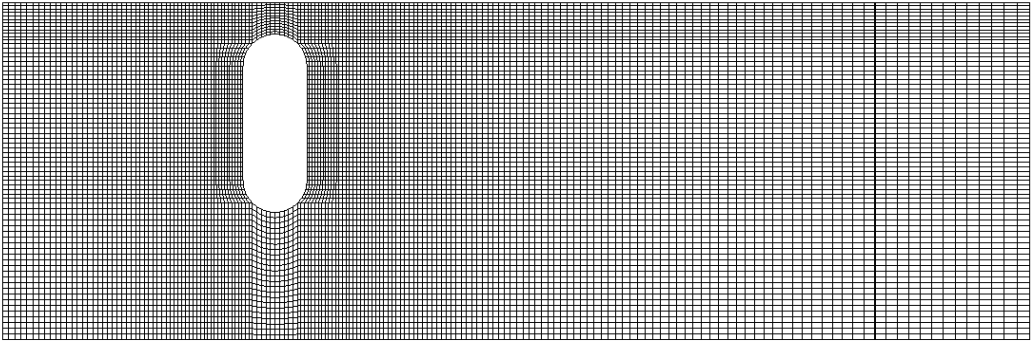}
\caption{2D flow lattice with a table shaped obstacle}
\label{2D flow lattice}
\end{figure}}
\newpage
{\begin{figure}[h]
\centering
\includegraphics[width=0.8\textwidth]{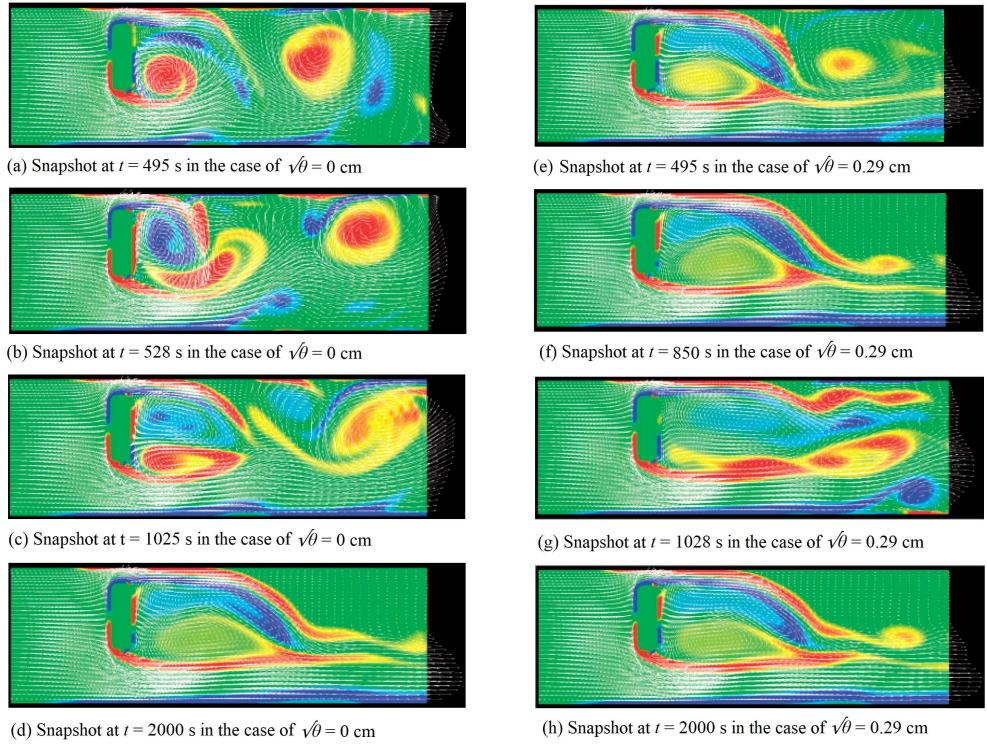}
\caption{(a)$\sim$(d)zero particle size case[movie 1], (e)$\sim$(h)non-zero particle size case[movie 2] } 
\label{Snapshots}
\end{figure}}

Let look at two movies 1 and 2
\footnote{\url{http://www2.yukawa.kyoto-u.ac.jp/~mayumi.saitou/simulation.html}}
:

1) The size of {\color{black}{a canal and an obstacle}}: 64[cm] (length) $\times$ 21[cm] (width) with small slit=2[cm] and large slit=8[cm]{\color{black}{, 
the obstacle size is 4[cm] (width) $\times$ 11[cm] (hight : including semicircles in both sides). 
Total number of grid point on the canal is 175 $\times$ 71, the minimum size of the grid length is 0.167[cm].}}

2) A time step is 0.001[s]; totally 2,000[s].  Reynolds number(Re)=700,
$\rho=10^3 [kg/m^3], \eta=8.85 \times 10^{-4} [Pa][s]$, and a typical velocity is $U_0=0.55 [cm/s]$.

3) Characteristics of the 1st  and the 2nd movies:

[movie 1] (commutative space with zero particle size) Before1000[s], vortices are
generated periodically; at 1025 [s] ``blue'' (anticlockwise
vortex) region and ``red'' (clockwise vortex)
region merges to form ``two attached eddies'';
at 2000[s] flow is stabilized.

[movie 2] (non-commutative space with non-zero particle size) No active periodic production of vortices just after the obstacle, but is
active at down stream; ``two attached eddies'' appear
from the beginning; after 1300[s] flow is stabilized.

\section{Quantization of Nambu's hydrodynamics (Part 2)}

It is important to note that Nambu's (hydro-)dynamics in D-dim phase space is equal to the dynamics of (D-2)-dim extended object (D-2)-brane; D=2: particle; D=3: string [Nambu(1980)]; D=4: membrane
[Su(1983)]; in general (Takhatajan(1994)] \cite{Action of Nambu dynamics} \footnote{This section started non-relativistically in the talk, so that there is a gap between this non-realtivistic treatment and the relativistic ones in the following sections.  To fill the gap, please follow Nambu and Sugamoto who studied relativistic hydrodynamics with 2-form and 3-form momenta, $p_{\mu\nu}$ and $p_{\mu\nu\rho}$. Then, the non-relativistic 3D acton Eq.(\ref{S3}) can be generalized relativistically as $S'_{3}=\int p_{\mu\nu}\; dX^{\mu} \wedge dX^{\nu} + \cdots$, including $\mu, \nu=0$.

In deriving the actions, they used the Hamilton-Jacobi (H-J) formalism.  The reason is as follows: The H-J is a method to give two actions, differing by total derivatives (or p-forms).  They are usually identical, giving the canonical transformation, but sometimes differ due to the non-integrable phase factors.  Therefore, Clebsch theory and gauge theories (with p-forms) in general fits to the H-J.}

For example, 3D hydrodynaics in (X, Y, Z) phase space can be described  by
\begin{eqnarray}
S_3= \int X dY \wedge dZ - (\psi_1 d\psi_2) \wedge dt. \label{S3}
\end{eqnarray}
Its equation of motion reads Nambu dynamics:
\begin{eqnarray}
\frac{\partial {\bf ~X}}{\partial t}= \{ {\bf X}, \psi_1, \psi_2 \}_{NB},~~ \mathrm{for} ~~ {\bf X}=(X, Y, Z).
\end{eqnarray}
This is a generalization of the usual 2D phase space $(x, p)$ case {\color{black}{of }}$S_2= \int p dx  - \psi_1 dt$.

Here, time t in 2D becomes ``area A" in 3D, namely,
\begin{eqnarray}
dx= \dot{x} dt, ~~  dY \wedge dZ = \frac{\partial(Y, Z)}{\partial(\sigma, t)}d\sigma \wedge dt 
 =\frac{\partial A^{23}(A) }{\partial A} dA.
 \end{eqnarray}

World sheet of string sweeps in (Y, Z), moving in t and parametrized by $(0 \le \sigma \le 2\pi)$; X is the momentum.

Path integration with weights, $e^{\frac{i}{\hbar}S_2}~\mathrm{and}~ e^{\frac{i}{\theta_3}S_3}$, gives quantum mechanics:

$\hat{p}= -i \hbar \frac{\partial}{\partial x}, ~ \mathrm{and} ~ \frac{\partial}{\partial x} x=1$
for 2D, while for 3D we have
\begin{eqnarray}
\hat{X}(\sigma)=\hat{X}^1(\sigma)= -i \theta_3 \frac{\delta}{\delta C^{23}(\sigma)}, \; \mathrm{and} \;  \frac{\delta}{\delta C^{23}(\sigma)} A^{23}(\sigma')=\delta(\sigma-\sigma').
\end{eqnarray}
The derivative in 3D is replaced by the functional derivative
(in tangential t and normal direction {\color{black}{$\mu$}}):
\begin{eqnarray}
\frac{\delta \Psi [C(\sigma)]}{\delta C^{t\mu}(\sigma)}= \lim_{\delta C(\sigma) \to 0} \frac{\Psi [C(\sigma)+\delta C^{t\mu}(\sigma)]-\Psi [C(\sigma)]}{\mbox{area of}~  \delta C^{t\mu}(\sigma)}.
\end{eqnarray}

Now, we have the uncertainty relation in 2D and 3D (see Appendix 2)):

\begin{eqnarray}
\sqrt{\langle (\Delta p)^2 \rangle} \sqrt{\langle (\Delta x)^2 \rangle} \geq \frac{\theta_2}{2} (=\hbar/2)~(D=2),~ 
\sqrt{\langle (\Delta X^1)^2 \rangle} \sqrt{\langle (\Delta A^{23})^2 \rangle} \geq \frac{\theta_3}{2}~(D=3),
\end{eqnarray}

Now, QED and its analog in 3D hydrodynamics read
\begin{eqnarray}
&&S_{\mathrm{particle}}= \int_{-\infty}^{\infty}dt \left\{ \left( p_{\mu} \frac{\partial x^{\mu}}{\partial t } +  g A_{\mu}(x) \frac{d x^{\mu}}{dt}\right) - \frac{1}{2}\left(\frac{1}{m}(p_{\mu})^2 +k (x^{\mu})^2\right) \right\}, \\
&&S_{\mathrm{string}}= \int_0^{2\pi} d\sigma \int_{-\infty}^{\infty}dt  \left\{  \sum_{XYZ:\mathrm{cyclic}}\left(X \frac{\partial (Y, Z)}{\partial(\sigma, t)} + g W(Y, Z) \frac{\partial (Y, Z)}{\partial(\sigma, t)}\right) \right. \nonumber \\
&&\hspace{4.5cm} \left. - \frac{1}{2} m(X^2+Y^2+Z^2)\right\}.
\end{eqnarray}
This is the Schild gauge string theory, coupled to Kalb-
Ramond field \footnote{A gap exists again between $S_{\mathrm{string}}$ and $S_{\mathrm{Schild}}$.  To fill the gap, please write the former action in terms of $X^{\mu}$ (including $\mu=0$}):
\begin{eqnarray}
S_{\mathrm{Schild}}=\frac{1}{2m} \int d\sigma dt \; \left(\frac{\partial(X^{\mu}, X^{\nu})}{\partial(\sigma, t)}\right)^2 + g W_{\mu\nu}(x)\frac{\partial(X^{\mu}, X^{\nu})}{\partial(\sigma, t)},
\end{eqnarray}
where $m$ can be $(\theta_3)^{1/3}$ and the string tension $\theta_3^{-2/3}$.

Our uncertainty relation to give the minimum size for a
granular particle is equal to the space-time uncertainty principle
claimed by Yoneya [Yoneya(1997)] \cite{Yoneya(1997)}.

$\bullet$ (i, 0) components of Kalb-Ramond field: The wave equation, describing the flow of fluid is
\begin{eqnarray}
i \theta_3 \frac{\partial}{\partial A} \Psi(C; A)= (\psi_1 d \psi_2) \Psi(C; A) 
\end{eqnarray}
which is equal to the ``gauge symmetry of string" in (i, 0) directions:
\begin{eqnarray}
\left\{i \theta_3 \; \frac{\delta \Psi(C; A)}{\delta C^{i0}(x)} - g W_{i0}(x) \right\} \Psi(C; A) =0.
\end{eqnarray}
Therefore, $gW_{i0}(x)= \psi_1 \frac{\partial}{\partial x^i} \psi_2$, and the vortex interaction is mediated by the stream functions, or equivalently by (i, 0) component Kalb-Ramond fields.

\section{A. Clebsch knew the Nambu dynamics in the middle of 19th century. [Clebsch (1857, 1859)]}

(see H. Lamb ``Hydrodynamics"\cite{Clebsch(1857-1859)}, a favorite book of Nambu.  He kept reading it until his death.)

Clebsch started with ${\bf v}=-{\bf \nabla}\phi(x) + \lambda(x) {\bf \nabla} \mu(x)$, then $\bm {\omega}= \bm{\nabla} \times \bm{v}=\{\bm{x}, \lambda, \mu \}_{NB}= {\bf \nabla}\lambda \times {\bf \nabla}\mu$; while Nambu started with ${\bf v}=\{{\bf x}, \psi_1, \psi_2\}_{NB}={\bf \nabla}\psi_1 \times {\bf \nabla}\psi_2$.

Nambu and Clebsch are ``dually" related, under 
$(v) \leftrightarrow (\omega)$, and (stream functions) $\leftrightarrow$ functions {\color{black}{$(\lambda,\ \mu)$}}.

Nambu studied the flow along velocity $(v)$, while Clebsch studied the flow along vorticity $(\omega)$.

Consider ``string=vortex line or tube", stretching in $\sigma$
direction, parallel to vorticity $\omega$,
\begin{eqnarray}
&\frac{\partial O}{\partial \sigma}=\; \{O, \lambda, \mu\}~ =(\omega \cdot \nabla) O~~~(\mathrm{Clebsch}), \\
&\frac{\partial O}{\partial t}=\{O, \psi_1, \psi_2\}=({\bf v} \cdot \nabla) O~~~(\mathrm{Nambu}).
\end{eqnarray}

Clebsch easily shows the vorticity theorem by Helmholtz, and also considered the reparametrization transformation $(\sigma, t) \rightarrow (\sigma', t')$.

This duality is probably ``T-duality" (translational mode
$\leftrightarrow$ winding mode) in string theory; suggesting a duality
in hydrodynamics.

\section{Hybrid model: vortex as string field, and the
hydrodynamics as local field}
This model was given 40 years ago by Nambu (1977)
and by one of the authors (A.S). [Su(1979), Seo-Okawa-Su{\color{black}{(1979,1981)}}] \cite{Hybrid}.

This Nambu's relativistic hydrodynamics be Kalb-Ramod
theory coupled with vorticity source: 
\begin{eqnarray}
L= \frac{1}{2}V_{\mu}V^{\mu}-\frac{k}{2} W_{\mu\nu}\omega^{\mu\nu}, ~~V^{\mu}=\frac{1}{2}\epsilon^{\mu\nu\lambda\rho}\partial_{\nu}W_{\lambda\rho},
\end{eqnarray}
where $W_{\mu\nu}$ is Kalb-Ramond field (1973) (now called B-field or anti-symmetric tensor field), and $V^{\mu}(x)$: fluid velocity, and $\omega^{\mu\nu}(x)$: vorticity source.  Then, the realtivistic fluid velocity satisfies the continuity relation naturally, $\sum_{\mu=0}^3\partial_{\mu} V^{\mu}=0$. 

$\bullet$ A filament of vortex line, extending in Z direction
with circulation $\gamma$, gives
\begin{eqnarray}
&\omega^{03}=(\nabla \times {\bf v})^3= \gamma \; \delta(x- X(t, \sigma)) \; \delta(y- Y(t, \sigma)) \\
& = \gamma \; dT(t, \sigma) \wedge dZ(t, \sigma) \; \delta^{(4)} (x^{\mu}- X^{\mu}(t, \sigma)),
\end{eqnarray}
so that
\begin{eqnarray}
\int dx \int d\sigma dt \; W_{\mu\nu}\omega^{\mu\nu}=\int d\sigma dt \gamma W_{\mu\nu}(x) \frac{\partial(X^{\mu}, X^{\nu})}{\partial(\sigma, t)}.
\end{eqnarray}

$\bullet$ Kalb-Ramond field is a ``gauge field" in string.
This gauge symmetry is called ``Kalb-Ramond symmetry":
\begin{eqnarray}
\Psi[C] &\rightarrow& \Psi'[C] = e^{i \oint_{C} dx^{t} \Lambda_{t}(x)} \Psi[C]  ,~\mathrm{and}\nonumber \\
W_{\mu\nu}(x) &\rightarrow& W_{\mu\nu}'(x) = W_{\mu\nu}(x) + \partial_{\mu} \Lambda_{\nu}(x) - \partial_{\nu} \Lambda_{\mu}(x) .
\end{eqnarray}

The hybrid model describes the interaction of vortex
strings through the fluid field (K-R field):
\begin{eqnarray}
\mathcal{L}_{\mathrm{hybrid}}& = \frac{1}{2} V^{\mu}(x)^2+\sum_{C} \left\{ \sum_{\mu \ne t} \left|  \left(-i \theta_3 \frac{\delta}{\delta C^{t\mu}(x)}- ig W_{t\mu}(x) \right) \Psi(C; A)\right|^2 \right\}. 
\end{eqnarray}

This is very similar to QED in which electric current couples to photon.
That is, in hydrodynamics, the interaction of vortices with fluid can be represented by  the coupling of current of string to K-R field:
\begin{eqnarray}
&&\mathcal{L}_{\mathrm{vortex-fluid}}=ig\theta_3^{2/3} \frac{1}{2} \sum_{\mu \ne t} W_{t\mu}(x) \left\{\Psi(C; A)^{\dagger} \left(\frac{\overleftrightarrow{\delta}}{\delta C^{t\mu}(x)}\right) \Psi(C; A) \right\} \nonumber \\
&&~~~~~~~~~~~~~~~+   \frac{g^2}{4\theta_3^{2/3}} \sum_{\mu \ne t} W_{t\mu}(x)W^{t \mu}(x) \left\{ \Psi(C; A)^{\dagger}\Psi(C; A)\right\} .
 \end{eqnarray}
Understanding of power spectrum of high energy cosmic ray (having Knee and Ankle)
can be applicable to power spectrum of vortices?

$\bullet$ An example of string field in the ordinary case: 
 Since $ \Psi[C, A] \approx \Psi[{\bf X}(\sigma, t)] $, the mode expansion $X^i(\sigma, t)=x^i+ \frac{\kappa}{\pi} (p^i t) +i \sqrt{\frac{\kappa}{4\pi}} \sum_{k\ne 0} \left( x^i_k e^{-ik(t-\sigma)} + \tilde{x}^i_k e^{-ik(t+\sigma)}\right), ~(\kappa=(\theta_3)^{2/3})$, gives the functional derivative 
\begin{eqnarray}
\frac{\delta}{\delta X^i(\sigma, t)}=2 \pi \frac{\partial}{\partial x^i}+\pi \left(i \sqrt{\frac{\kappa}{4\pi}}\right)^{-1} \sum_{k\ne 0} \left( e^{+ik(t-\sigma)} \frac{\partial}{\partial x^i_k} +  e^{+ik(t+\sigma)}\frac{\partial}{\partial \tilde{x}^i_k} \right).
\end{eqnarray}

Then, the wave equation reads
\begin{eqnarray}
&\left[ \partial_{\mu}\partial^{\mu} + \frac{1}{2}\sum_{i} \sum_{k \ne 0}\left( \frac{\partial^2}{\partial x^i_k \partial x^i_{-k}} + k^2 x^i_k x^i_{-k} \right) \right] \Psi[x^{\mu}, \{x^i_k \}]\\
&=\left[ -p^2 +\left\{ \left( \sum_i \sum_{k=1}^{\infty} k(n^i_k + \tilde{n}^i_k)\right) -2 \right\} \right]\Psi[x^{\mu}, \{x^i_k \}].
\end{eqnarray}

$\bullet$ String has, in addition to the translational mode,
the k-th oscillation mode in the i-th direction with
occupation number $\{n^i_k\}$.
The wave functional becomes
\begin{eqnarray}
\Psi[X(\sigma, t)] \propto  \sum_p \sum_{\{n^i_k\}} \left\{ \hat{A}_{p, \{n^i_k\}} e^{-i px} \prod_i \prod_{k=1}^{\infty} H_{\{n^i_k\}}(x^i_k)e^{-k (x^i_k)^2}+ \hat{\tilde{A}}_{p, \{\tilde{n}^i_k\}}\times [ \{n^i_k\} \to \{\tilde{n}^i_k\} ] + (h.c.) \right\},
\end{eqnarray}
where $\hat{A}_{p, \{n^i_k\}}$ and  $\hat{\tilde{A}}_{p, \{\tilde{n}^i_k\}}$ are annihilation operator of string itself of right- and left-moving, respectively.
Here, we have to choose a proper string theory (not necessary to follow the usual examples) so as to reproduce the vortex production rate in a given hydrodynamics.

\section{Distribution function of vortices inside turbulence}
A nucleon is composed of quarks and gluons (partons);
having momentum fraction x relative to P of the nucleon.
Patrons are probed by photon {\color{black}{$\gamma$}}, with momentum Q. 
If $Q^2$ is increased, the finer structure reveals:
{\begin{figure}[h]
\centering
\includegraphics[width=0.8\textwidth]{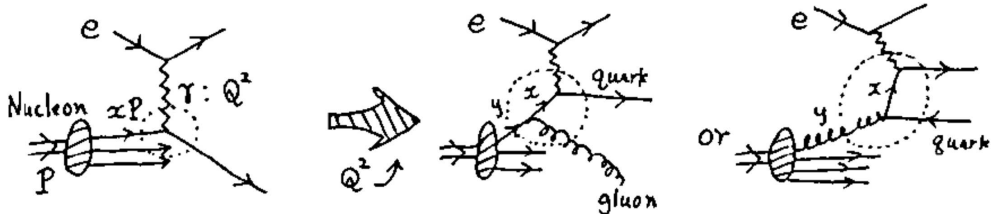}
\caption{Evolution of quark's distribution function in QCD}
\label{QCDevolution}
\end{figure}}\\
Similarly, a turbulence is composed of vortices (partons)
with various sizes and shapes, \\
{\begin{figure}[h]
\centering
\includegraphics[width=0.8\textwidth]{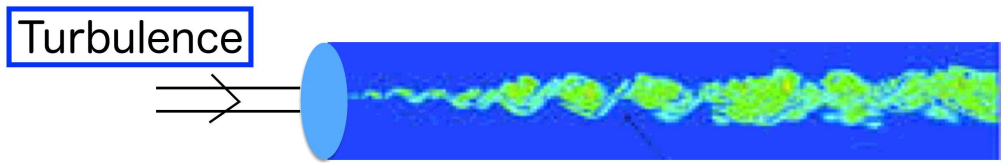}
\caption{Distribution function of turbulence}
\label{Turbulence}
\end{figure}}
\newpage\noindent
A parent turbulence (like quark) moves in z direction ({\it i.e.} in the frame where the observer moves in -z direction very {\color{black}{fast)}}, the momentum in z
direction is dominantly large, but the other transverse momenta in (x, y) directions are small. Then, the vortex (daughter patrons) inside a parent turbulence (with momentum $\bm{P}$), has the momentum ${\bf p}= x{\bf P}^+ + {\bf p}^{\bot},\; ({\bf p}^{\bot})^2=Q^2$.
{\begin{figure}[h]
\centering
\includegraphics[width=0.8\textwidth]{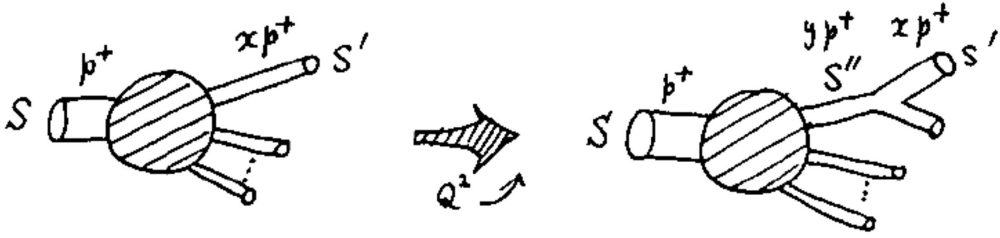}
\caption{Evolution of closed string (vortex)}
\label{Stringevolution}
\end{figure}}

The evolution for closed string (vortex) was given 30 years
ago by us [MaruiOdaSu(1990)], in a similar manner as the Altarelli and Parisi (1973) equation in QCD \cite{AP}.

Distribution function $D_{S'S} (x, Q^2)$ of string $S'$ inside $S'$ 
with momentum fraction $x$ at the energy scale $Q^2$ satisfies
\begin{eqnarray}
Q^2 \frac{d D_{S'S}}{d Q^2} (x, Q^2) = \sum_{S''}\int_0^1 \frac{dy}{y} P_{S'S''}\left(\frac{x}{y}, Q^2 \right) D_{S''S} (x, Q^2) \nonumber \\
- \sum_{S''} \int_0^1 dz P_{S'S''}\left(z, Q^2\right) D_{S''S} (x, Q^2),
\end{eqnarray}
where the splitting function is given in Figure 6.
{\begin{figure}[h]
\centering
\includegraphics[width=0.8\textwidth]{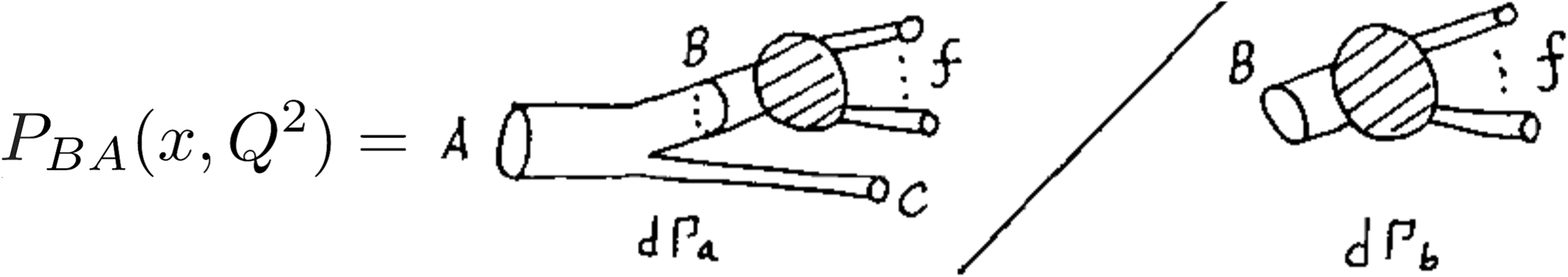}
\caption{Splitting function of string (vortex)}
\label{splitting}
\end{figure}}

Using the light-cone field theory of string by Mandelstam
(1973) and Kaku-Kikkawa (1974) \cite{string field theory}, we obtain
\begin{eqnarray}
P_{BA}(z, Q^2) \sim g^2 e^{Q^2 F_{BA}(z)}, \; F_{BA}(z)=-\frac{z \ln z +(1-z) \ln (1-z)}{z(1-z)}
\end{eqnarray}
which differs from the usual distribution functions of
quarks and gluons,
\begin{eqnarray}
P_{qq}(z, Q^2) \sim g_s^2(Q^2) \left(\frac{1+z}{1-z}\right), \; P_{gq}(z, Q^2) \sim g_s^2 (Q^2)\left(\frac{1+(1-z)^2}{z}\right).
\end{eqnarray}

$\star$ Note: The concept itself is applicable to the
turbulence or other vortex phenomena,
but we have to choose the more proper string theory
(not necessary the usual ones), so that it may reproduce the
realistic phenomena in a given hydrodynamics.

\section{(Simple-minded) inclusion of shear viscosity}
Shear viscosity $\eta, ~(\eta/\rho=\nu)$ introduces diffusion in
velocity field: $\dot{\bm{v}}= \Delta \bm{v}$~(constitutive equation).
To include the thermal fluctuations at finite temperature
the ``constitutive equation'' is relaxed by the
fluctuation of force, following the Gaussian distribution
with width proportional to $T$:
\begin{eqnarray}
\dot{\bm{v}}- \nu \Delta \bm{v} = \bm{\xi(t)},~~\mathrm{with}~Pr [\bm{\xi}] \propto  e^{-\frac{\alpha}{2k_BT} \int dt (\bm{\xi}(t))^2}.
\end{eqnarray}

In thermodynamics, $\xi$ plays the same role as the
momentum $p$ in quantum mechanics, and hence
\begin{eqnarray}
&&L^{\star}= \left\{\dot{\bm{x}} \cdot \left( \bm{p} - g \bm{A}(x)\right) - \left( \frac{1}{2m} \bm{p}^2 + \frac{k}{2} \bm{x}^2 \right) \right\} + i  \left\{ \bm{\xi} \cdot (\dot{\bm{v}} -\nu \Delta \bm{v})- \frac{\alpha'}{2k_BT} (\bm{\xi})^2 \right\} \nonumber \\
&&~~~~~+ \bm{\lambda}(t) (\dot{\bm{x}}-\bm{v}).
\end{eqnarray}

We can include the dissipation effects also in string
theory, following for example [Aibara et al.(2019)] \cite{OUJ}.
In the particle theory we have
\begin{eqnarray}
&&i \hbar  \frac{\partial}{\partial t} \psi(\bm{x}, t) = \left[ \left\{ \frac{1}{2m} \left(- i\hbar\bm{\nabla}_x+ g \bm{A}(\bm{x})\right)^2 +\frac{1}{2}k \bm{x}^2  \right\}  \right.\nonumber \\
&&~~~~~~~~~\left. +i \left\{ 2\alpha'k_B T \left(  \bm{\nabla}_v + \frac{1}{2\alpha'} \nu \Delta \bm{v} \right)^2 - \frac{k_B T}{2\alpha'}  \nu^2 ( \Delta \bm{v} )^2 \right\} \right] \psi(\bm{x}, t).
\end{eqnarray}
under the constraint of $\dot{\bm{x}}=\bm{v}$, or
$\left(- i\hbar\bm{\nabla}_x+ g \bm{A}(\bm{x})-m\bm{v} \right) \psi(\bm{x}, \bm{v}, t) =0.$

The corresponding equation in string is
\begin{eqnarray}
&& i \theta_3 \;  \partial_A \Psi(C, D; A)=\left[ \frac{1}{2\theta_3^{1/3}} \left\{ \sum_{\mu \ne t}  \left(-i \theta_3 \frac{\delta}{\delta C^{t\mu}(x)}- g W_{t\mu}(x) \right)^2  \right\} \right. \nonumber \\
 && \left. + i k_BT \left\{ 2\alpha'  \left(\frac{\delta}{\delta D^{t\mu}(x)}+ \frac{1}{2\alpha'}\nu \Delta_x D_{t\mu} (x) \right)^2 - \frac{1}{2\alpha'} \nu^2 (\Delta_x D_{t\mu}(x))^2 \right\} \right] \Psi(C, D; A),~~~
 \end{eqnarray}
 with a constraint, representing 
 \begin{eqnarray}
 \left\{-i \theta_3 \frac{\delta}{\delta C^{t\mu}(x)} -  D_{t\mu}(x) \right\} \Psi(C, D; A)=0,
 \end{eqnarray} 
where $\frac{\partial C^{t\mu}}{\partial A}(x)=D^{t\mu}(x)$, and $C^{t\mu}=X^t \wedge X^{\mu}$. 

\section{Is Landau's (relativistic) fluid dynamics persisted or modified?\cite{Landau}}

4D Energy momentum tensor in particle thery $T^{\mu\nu}(x)=\frac{\partial\mathcal{L}}{\partial(\partial_{\mu}\phi(x))}\partial^{\nu}\phi(x) - g^{\mu\nu}(x) \mathcal{L}$ becomes in string
\begin{eqnarray}
&&T^{\mu\nu}[C]=\Psi[C]^{\dagger} \left\{ \frac{\delta^2}{\delta X_{\lambda}[C] \delta X^{\lambda}[C] } - \frac{\alpha}{(\theta_3)^{4/3}} \partial^{\nu} \partial_{\sigma}X_{\mu}[C]X^{\mu}[C] \right. \nonumber \\
&&\left. - g^{\mu\nu}(X) \left( \frac{\delta^2}{\delta X_{\lambda}[C] \delta X^{\lambda}[C] } - \frac{\alpha}{(\theta_3)^{4/3}} \partial^{\nu} \partial_{\sigma}X_{\mu}[C]X^{\mu}[C] \right) \right\}\Psi[C].
\end{eqnarray}

Oscillations give various ``strain" to string
\begin{eqnarray}
T^{\mu\nu}[C]=\Psi(x^{\mu}; \{x_k^i\}) \left\{ t^{\mu\nu} \left(x^{\mu}, \frac{\partial}{\partial x^{\mu}} ; \{x_k^i\}, \frac{\partial}{\partial \{x_k^i\}} \right) \right\}\Psi(x^{\mu}; \{x_k^i\}).
\end{eqnarray}
Therefore, the derivative expansion is surely modified ?
\begin{eqnarray}
&& t^{\mu\nu}=a_0(x)+a_2(x) \partial^{\mu}\partial^{\nu}+b_2(x) \sigma_{k\ne 0} \left(\partial^{\mu}\frac{\partial}{\partial x_k^{\nu}} +\partial^{\nu}\frac{\partial}{\partial x_k^{\mu}} \right) + c_2(x) \partial^{\mu}\partial^{\nu}\partial^{\nu}\frac{\partial}{\partial x_k^{\mu}} + \cdots, ~~\mathrm{or} 
\quad\quad\quad\\
&&\langle t^{\mu\nu}\rangle= p(x) g^{\mu\nu}(x) (\mathrm{pressure}) + \rho(x) v^{\mu}v^{\nu} (\mathrm{advective~term}) \nonumber \\
&&~~~~~~~+ \eta(x) \sum_{k\ne 0} \left( \partial^{\mu} \varepsilon^{\nu}_k +\partial^{\nu} \varepsilon^{\mu}_k \right) (\mathrm{viscosity~by~strain} \varepsilon^{\nu}_k) + \cdots.
 \end{eqnarray}
 
 \section{Conclusions}
\begin{enumerate}
\item[(1)]Nambu's hydrodynamics in non-commutative space
can be the hydrodynamics of granular material.
\item[(2)]The non-commutativity (or quantization) of space can be
introduced, by 1) Moyal product, 2) vortex as string and its
coupling to Kalb-Ramond field, or 3) string field theory.
\item[(3)]In applying the string field theory,
1) evolution equation describing the change of energy
scale, and 2) Landau's way of derivative expansion of
energy momentum tensor, can be applied.
\item[(4)]Note however: We have to find a proper string
theory so as to reproduce a given hydrodynamics.
\item[(5)]Force from the stress tensor should be re-examined.
\end{enumerate}
\section*{Acknowledgements}
The authors give their sincere thanks to Professor Yasuhide Fukumoto and Professor Hiroshi Itoyama for inviting one of them (A.S.) to the workshop, where very active discussions were exchanged between the participants.  Especially, we are grateful to Yutaka Matsuo for clarifying a number of unclear points in their talk during and after the workshop, which triggers a new insight by Yutaka Matsuo and A.S. to appear in the proceedings. 

This work was partly supported by Osaka City University Advanced Mathematical Institute (MEXT Joint Usage/Research Center on Mathematics and Theoretical Physics JPMXP0619217849). 
{\color{black}{Numerical computation in this work was carried out at the 
Yukawa Institute Computer Facility.}}

=============================================
\section*{Appendices:~Reserved Slides}
\section*{$\bullet$ Appendix 1): Hydrodynamic simulation}
(A1-1) : The fractional step method with a tentative $v*$ is used:
\begin{eqnarray}
v_i^{*}&=&v_i^{(n)} + \frac{\Delta t}{\rho}
\left[\ -(\bm{v}\cdot \nabla) v_i + \eta \Delta v_i+K_i\ \right]^{(n)}.
\end{eqnarray}
Applying the continuity condition for $v_i^{(n+1)}$, $p$ can be
determined by ${\color{black}{\Delta p}} =\frac{\rho}{\Delta t}\ \nabla\cdot \bm{v}^{*}$. 

(A1-2): ``The 3rd order upwind difference" is used for the
advective term.

(A1-3): About the boundary conditions, since the differential equations are the 3rd order ones, we impose, 1) $\bm{v}=\bm{0}, ~(\bm{\nabla}p)_{\bot}=\bm{0}, ~\bm{K}_{\bot}=\bm{0}$ on the boundary,  2) $(\bm{\nabla}p)_{\bot}=\bm{0}$ at the entrance, and 3) $p=0$ at the exit.

\section*{$\bullet$ Appendix 2): Every operator algebra is imprinted
in the path integral expression} 
See Feynman's paper (1948), and\\
{\begin{figure}[h]
\centering
\includegraphics[width=0.6\textwidth]{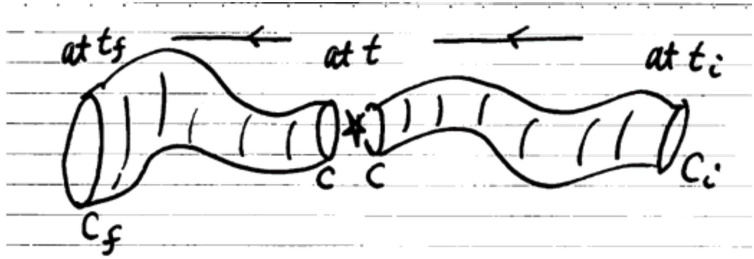}
\caption{A method to obtain operator algebra in path integral}
\label{Feynmann}
\end{figure}}\\
insert $\frac{\partial O(\bm{X}(\sigma)}{\partial t}$ and $\frac{\delta A^{23}[C(\sigma)]}{\delta C(\sigma)}$ at the star, and perform
integration by parts.  Then we have
$i \theta_3 \; \frac{\partial O(\bm{X}(\sigma)}{\partial t} = [O(\bm{X}(\sigma), \oint \psi_1 d \psi_2]$, and $i \theta_3 \delta(\sigma-\sigma')=[A^{23}(\sigma, t), X^1((\sigma', t)]$, 
in the same manner as Feynman did for
$\frac{d O(t)}{dt} ~\mathrm{and}~ \frac{dx(t)}{dx}$. 

\section*{$\bullet$ Appendix 3): Interaction between vortices}
Without viscosity $\bm{v}=\bm{\nabla} \times \bm{\psi}$, we have a constant $\bm{\nabla} \cdot \bm{\psi}$, and $\bm{\omega}=\bm{\nabla} \times \bm{v}=-\Delta \bm{\psi}$.  Then, in hydrodynamics 
\begin{eqnarray}
E_{\mathrm{Bernoulli}}=\frac{1}{2} \int d^3 x \; \rho \bm{v}^2 = - \frac{1}{2} \int d^3x \;  \rho \; \bm{\psi} \Delta \bm{\psi} = - \frac{1}{2} \int d^3 x \int d^3 y\; \rho \; \bm{\omega}(x) D(x-y) \bm{\omega}(y),
\end{eqnarray}
while in QED: 
\begin{eqnarray}
&&E_{QED}=\frac{1}{2} \int d^3 x \int d^3y  ( \bm{E}^2+\bm{B}^2) \\
&&= - \frac{1}{2} \int d^3x \int d^3y \left\{ \frac{1}{4 \pi \epsilon_0}\; q(x) D(x-y) q(y) - 4 \pi \mu_0 \; \bm{j}(x) D(x-y) \bm{j}(y) \right\}.  
\end{eqnarray} 

Now we can understand that attraction and repulsion is the same for electric
charge and vortex, but is reversal for electric current.

In the hybrid model of hydrodynamics, vortex interaction is mediated by the Kalb-Ramond field: 
\begin{eqnarray}
&&V^{\mu}=\frac{1}{2}\epsilon^{\mu\nu\lambda\rho}\partial_{\nu}W_{\lambda\rho}=\frac{1}{3!} \epsilon^{\mu\nu\lambda\rho}(\partial_{\nu}W_{\lambda\rho}+\partial_{\lambda}W_{\rho\nu}+\partial_{\rho}W_{\nu\lambda}) \nonumber \\
&&\equiv \frac{1}{3!} \epsilon^{\mu\nu\lambda\rho}W_{\nu\lambda\rho}, 
\end{eqnarray}
with the components
\begin{eqnarray}
&&e^i= W^{0i}, \; b^i=\frac{1}{2} \epsilon_{ijk} W^{jk}, \\
&&V_0=V^0=\bm{\nabla}\cdot \bm{b}
, \; V_i=V^{i}=[ \dot{\bm{b}}+ \bm{\nabla} \times \bm{e}]^i. \label{KR-velocity}
\end{eqnarray}

Now, the vortex interaction is correctly reproduced:
\begin{eqnarray}
&S= \int dx\left\{ \frac{1}{2} (V^{\mu})^2-\frac{1}{2}g W_{\mu\nu} \omega^{\mu\nu}\right\}= \int dx \frac{1}{2} \left\{ (\nabla\cdot{\bf b})^2 
- (\dot{\bf b}+ \nabla \times {\bf e})^2 -g (e^i \omega^{0i}) \right\}\\
& ~~\to  -\frac{1}{2}\int d^3x \int d^3y \; \sum_{i=1}^3 \omega^{0i}(x) D(x-y) \omega^{0i}(y).
\end{eqnarray}

The vortex interaction is mediated by $e(x)$ which are
the Kalb-Ramond fields or the stream functions: $e^i(x)=W^{0i}=\psi^i(x)$.

$\bullet$ As we know that the massless modes of closed string in our simple case are dilation (D), graviton (G) and Kalb-Ramond (W) fields.
\begin{eqnarray}
\Psi [X(\sigma)]|_{\mathrm{massless}}= D(x) \eta_{\mu\nu}\alpha_{-1}^{\mu} \tilde{\alpha}_{-1}^{\nu} + \frac{1}{2} G_{\mu\nu}(x) (\alpha_{-1}^{\mu} \tilde{\alpha}_{-1}^{\nu}+\alpha_{-1}^{\nu} \tilde{\alpha}_{-1}^{\mu}) +\frac{1}{2} W_{\mu\nu}(x) (\alpha_{-1}^{\mu} \tilde{\alpha}_{-1}^{\nu}-\alpha_{-1}^{\nu} \tilde{\alpha}_{-1}^{\mu}).
\end{eqnarray}
in the first quantized notation, that is,
\begin{eqnarray}
\hat{A}^{\dagger}_{p, \{n_k^i\}}|0\rangle\rangle_{\mathrm{2nd~quantization}}=\prod_{i \bot t} \; \prod_{k=1}^{\infty} \alpha_{-k}^{i} | p \rangle_{\mathrm{1st~quantization}}.
\end{eqnarray}

\section*{$\bullet$ Appendix 4): Different shape of granular particle}
{\color{black}{To differentiate the shape of particle, }}
1) consider the particle as a rigid body, and
2) impose the symmetry.
For 1), introduce the body fixed frame and Euler angles, and specify
the configuration of the frame, relative to the space fixed frame.
For 2), impose the invariance under the ``point group G" of the
particle; SO(3) for a sphere, tetrahedral group T for a tetrahedron,
and octahedral group O for a cube.
The flow of six variables (velocities and Euler angles), or the flow
of four stream functions, being invariant under the point group
G; the additional two stream functions are Hamiltonian and
angular momentum squared of the fluid particle as rigid body,
which also follows the Nambu dynamics.

\section*{$\bullet$ Appendix 5): In granular hydrodynamics, energymomentum
tensor takes a completely different form ?}

The usual form $T^{ij}=-p \delta^{ij} + \eta \left( \partial^i v^j + \partial^j v^i -\frac{2}{3} (\nabla {\bf v}) \right) + \zeta \delta^{ij} (\nabla {\bf v})$ is surely violated.
Can we modify it and understand the following
phenomena in granular hydrodynamics?

$\bullet$ (Open Problem): Various oscillation modes in string give
various types of strain; is this useful to understand the
above phenomena?  For example, 
1) flow of sands in hourglass, 2) arc structure of
sands clogging in a funnel, 3)
Reynolds ``dilatancy": decrease of density by the increase of
strain, 4) Brazilian-nuts effect {\it e.t.c.} (see \cite{Dulan}).

$\bullet$ (Open Problem): Various oscillation modes in string give
various types of strain; is this useful to understand the
above phenomena ?

 



\begin{thebibliography}{99}

\bibitem{Nambu1973} \underline{Nambu dymamics}:  Y. Nambu, Phys. Rev. {\bf D7} (1973) 2405. \\

\bibitem{Nambu(2011-2013)} \underline{Nambu's last work}:  Y. Nambu, Talk at International Workshop: Extra Dimensions \\
in the Era of the LHC, Dec. 12-14 (2011). 
\url{https://sites.google.com/a/hetmail.phys.sci.osaka-u.ac.jp/edlhc11/};\\ 
Y. Nambu, Talk at Int. Symposium on Research Frontiers on Physics, Earth and Space Science, Dec. 17-18 (2013). \url{http://thmat8.ess.sci.osaka-u.ac.jp/Meeting2013/Talks/YN.pdf} \\

\bibitem{SaBSu(2014)} \underline{Hydrodynamics of granular material}: M. Saitou, K. Bamba and A. Sugamoto: PTEP 103B03 (2014); M. Saitou: phD thesis, Ochanomizu U., March (2016).\\

\bibitem{KaKuNaSaSu(2018)}\underline{Hydrodynamic simulation}: T. Kawamura, A. Kuwana, Y. Nagata, M. Saitou and A. Sugamoto, PTEP  063J01(2018); 
{\color{black}{T. Kawamura, A. Kuwana, M. Saitou, A. Sugamoto, Y. Nagata,
RIMS Kokyuroku (2019), 2117:35-43. \url{http://hdl.handle.net/2433/252120} }} \\

\bibitem{Action of Nambu dynamics} \underline{Action of Nambu dynamics}: Y. Nambu: Phys.\ Lett.\ {\bf 92B}, 327 (1980); A. Sugamoto: Nucl. Phys. {\bf B215}, 381 (1983); 
L.~Takhtajan: Commun.\ Math.\ Phys.\  {\bf 160}, 295 (1994). \\

\bibitem{Yoneya(1997)} \underline{Space time uncertainty principle}: T. Yoneya, PTP {\bf 97}, 949 (1997).\\

\bibitem{Clebsch(1857-1859)}
\underline{Clebsch's work (1857,1859)}: referred 
in ``Hydrodynamics'' by Horace Lamb, Cambridge U. Press (1916). \\

\bibitem{Hybrid} \underline{Hybrid model of hydrodynamics}: M. Kalb and P. Ramond, Phys. Rev. {\bf D9 } (1974) 2273; 
Y.  Nambu: ``Quark Confinement and Field Theory", eds. D. R. Stump and D. H. Weingarten, John Wiley and Sons (1977) 1; 
A. Sugamoto: Phys. Rev. {\bf D19} (1979) 1820; K. Seo and A. Sugamoto: Phys. Rev. {\bf D19} (1979) 3744; {\it i.b.d.} {\bf D24} (1981) 1630; A. Sugamoto, Lecture at Kashikojima Summer Institute, ``Old Fashioned Duality Revisited'', arXiv: hep-th/9611051 (1996);
R. Endo, K. Seo, A. Sugamoto, Grav. Cosmol. {\bf 8} (2002) 56.\\

\bibitem{AP} \underline{Evolution equation of QCD and string}: (QCD) G. Altarelli and G. Parisi, Nucl. Phys. {\bf B126} (1977) 298; (string) M. Marui, I. Oda, and A. Sugamoto, Int. J. Mod. Phys. {\bf A5} (1990) 4257; A. Sugamoto, Proceedings of the Trieste conference on ``Supermembrane and Physics in 2+1 Dimensions'' (1989).\\

\bibitem{string field theory} \underline{String field theory}:  S. Mandelstam, Nucl. Phys. {\bf B64} (1973) 205; {\it i.b.d.}{\bf B69} (1974) 77;   M. Kaku and K. Kikkawa, Phys. Rev. {\bf D10} (1974) 1110, 1823.\\

\bibitem{OUJ} \underline{Non-equilibrium thermodynamics}: N. Aibara, N. Fujimoto, S. Katagiri, M. Saitou, A. Sugamoto, T. Yamamoto, and T. Yumibayashi, PTEP 073A02 (2019).\\

\bibitem{Landau} \underline{Landau's fluid dynamics}: Sec.XV ``Relativistic Fluid Dynamics'' in ``Fluid Mechanics'' by L. Landau and E. M. Lifshitz.

\bibitem{Feynmn}
R.~P.~Feynman, Rev.\ Mod.\ Phys. {\bf 20}, 367 (1948).

\bibitem{Dulan}
J. Dulan, ``Physics of granuler material'' (1997) Editoins Eyrolles, Paris, translated in Japanese by Shu Nakanishi and Ko Okumura, (2001) Yoshioka-Shoten. 


\end{thebibliography}
\end{document}